\documentclass[prb,aps,floatfix]{revtex4}
\usepackage{graphicx}
\usepackage{tikz}

\begin{document}
\title{Physics-informed machine learning for the COVID-19 pandemic: Adherence to social distancing and short-term predictions for eight countries.}
\author{G. D. Barmparis and G. P. Tsironis}
\affiliation{
Institute of Theoretical and Computational Physics and Department of Physics, University of Crete, P.O. Box 2208, 71003 Heraklion, Greece\\
}
\date{\today}

\begin{abstract}
The spread of COVID-19 during the initial phase of the first half of 2020 was curtailed to a larger or lesser extent through measures of social distancing imposed by most countries. In this work, we link directly, through machine learning techniques, infection data at a country level to a single number that signifies social distancing effectiveness. We assume that the standard SIR model gives a reasonable description of the dynamics of spreading, and thus the social distancing aspect can be modeled through time-dependent infection rates that are imposed externally.
We use an exponential ansatz to analyze the SIR model, find an exact solution for the time-independent infection rate, and derive a simple first-order differential equation for the time-dependent infection rate as a function of the infected population. Using infected number data from the ``first wave" of the infection from eight countries, and through physics-informed machine learning, we extract the degree of linear dependence in social distancing that led to the specific infections. We find that in the two extremes are Greece, with the highest decay slope on one side, and the US on the other with a practically flat ``decay". The hierarchy of slopes is compatible with the effectiveness of the pandemic containment in each country. Finally, we train our network with data after the end of the analyzed period, and we make week-long predictions for the current phase of the infection that appear to be very close to the actual infection values.

\end{abstract}

\maketitle

\section{Introduction}
The COVID-19 pandemic started in December 2019 in China and subsequently spread fast in the rest of the world. After an initial ``hesitant" approach, most countries essentially adopted rules of social distancing that also originated in China. Several countries delayed the imposition of measures and, as a result, saw large numbers of infected persons and deaths. Other countries acted very swiftly and managed to control the infected numbers and especially the mode of spreading. There was an initial discussion related to ``herd immunity" that was in part attempted by some countries, but soon the basic global approach was that introduced by China, i.e. social distancing. However, the degree and swiftness of social distancing were different in each country; in Italy and Spain, for instance, there was an initial delay while Greece acted very quickly and with strong measures. Although ultimately the effectiveness of any measure is reflected in the number of deaths, in this work, we use the more error-prone infection data for a number of reasons. The infection data are representative of the dynamics of the disease at the country level, even though they clearly depend on the number of tests performed. At the initial phase, the test availability was limited, and thus it was used on a need to be the basis and, as a result, targeted more closely individuals with symptoms.  Additionally, since the COVID-19 pandemic affects people of older ages primarily, the death data are strongly age biased and thus do not reflect the true dynamics of the spreading that leads to these deaths.

In an earlier publication, the first version of which appeared in the arXiv on March 31, 2020, i.e. right in the middle of the pandemic, we used a Gaussian hypothesis for the spreading of the disease and number of infected persons and predicted the spreading and the horizon of the first wave \cite{BT}.  We showed that this specific functional dependence originated from the imposed measures and, in particular, from an approximately linear reduction in the infection rate $\alpha (t)$ as a result of imposed measures. This hypothesis proved to have two-fold usefulness: On one hand, it gave a good prediction for the horizon of the epidemic in countries like Greece, Italy, and Spain while the measures were in effect. On the other hand, for countries such as the US and UK where measures either did not enforce in full strength or were not applied fast enough, the prediction of the model based on the Gaussian hypothesis was rather poor. Although this was expected, it nevertheless gives a very good way to assess now, i.e. after the fact, how efficient were the measures in these and other countries. This may be done by evaluating from the real data an effective number that gives a degree of the harshness of imposed measures, adequate timing, etc.  This number, denoted by  $\sigma$, is the slope of the assumed linear dependent decay of the infection rate coefficient. Large $\sigma$ means that the effective measures where drastic and applied on time while, in the other extreme, $\sigma \approx 0 $ signifies the practical absence of measures.

In order to perform the present study, we use the following approach: We start with an SIR model \cite{KM} and use analytics in order to derive a differential equation of the infection rate $\alpha (t)$; this equation contains the information on the individual infection percentage in the population. We then take the data for the country's infected population and estimate the infection rate $\alpha (t)$. This step is done by using Machine Learning (ML) techniques and, in particular, by using physics-informed neural networks (PINN). We pre-train the latter on simulated SIR data and subsequently train it on each country's reported infected data. Instead of assessing a general $\alpha (t)$ curve, we assume a linear functional dependence explicitly; its slope $\sigma$ is the result of the ML procedure we apply. Once the infection rate is known, we validate the resulting SIR model to the country's data and then vary sigma to see the changes in the epidemic. This procedure gives a clear picture of both of the effective measures in each country but also their efficiency.

The assumption of linear decay in $\alpha (t)$ with slope $\sigma$ is tantamount to an effective linearization to the actual infection rates.  Clearly, other, more complex forms may be assumed. We find that this simple form can efficiently capture the nature of the phenomenon and give a simple quantitative estimate of the imposed measures.  The values of the slope $\sigma$ are obtained directly through ML and, thus, in a sense, are directly derived from the infection data. Thus, we may link each infection curve with an effective decay slope $\sigma$ that denotes the overall control that the measures exercise on the infection phenomenon.  Since the approach is fundamentally data-driven, the knowledge of a particular slope gives a handling on the possible measures exercised.  Furthermore, once the PINN we develop works well, we may use it to make predictions. Specifically, we use data from the second phase of the spreading, that we assume starts after the initial decline in the infection, train the network with this data and make short term predictions for the current period.

The structure of the paper is thus the following. In the next section II, we map the SIR system of two first-order ODE's into a unique second-order one that, in the case of constant $\alpha (t)$, may be solved exactly.  Subsequently, in section III, we derive a first-order equation for $\alpha (t)$ and write its general solution of arbitrary time dependence in the infected population. We verify explicitly that when the population dynamics is Gaussian, the dominant decay in $\alpha (t)$ is linear, as discussed in a more restricted form in \cite{BT}. Subsequently, in section IV, we apply our ML arsenal to derive $\alpha (t)$ from each country's data and determine the spectrum of $\sigma$'s for eight countries. We use this information back to the SIR model to investigate the actual as well as other hypothetical scenarios for the evolution of the epidemic in each country. This step gives a clear picture of the effectiveness of measures in each country. In section V, we use the data from the first phase of the pandemic for the eight countries, train the PINN with this data except for the last week of available data and subsequently make predictions for the evolution during that week and compare with the available data.  Finally, in section VI, we summarize our findings and conclusions. In Appendix A, we present the solution of the SIR model obtained through the exponential ansatz as well as approximate solutions in extreme limits to demonstrate the infection's behavior.  In Appendix B, we present a flowchart that details the ML approach used in this work.

\section{Exact mathematics of SIR}
The simple Susceptible-Infected-Removed (SIR) infection model is very powerful in 
determining qualitative but also quantitative aspects of the COVID-19 pandemic\cite{KM}. The basic equations are
\begin{eqnarray}
\label{Eq-1a}
\frac{dS}{dt} = - \alpha  S I\\
\label{Eq-1b}
\frac{dI}{dt} = \alpha S I - \mu I
\end{eqnarray}
where $S \equiv S(t) $, $I \equiv I (t)$ are the percentage of susceptible and infected individuals  respectively and the infection and removal rates $\alpha \equiv \alpha (t)$, $\mu = \mu (t)$ respectively
are functions of time in general. We introduce the variable $q(t)$ through the ansatz:
\begin{equation}
\label{Eq-2}
I(t) = e^{q(t) - \int_0^t \mu (t') dt'}
\end{equation}
Upon substitution to the set of Eqs (\ref{Eq-1a},\ref{Eq-1b}) we obtain
\begin{eqnarray}
\label{Eq-3a}
\dot{S} = - \alpha  e^{q-\nu}S \\
\label{Eq-3b}
\dot{q}  = \alpha S \\
\label{Eq-3c}
\nu \equiv  \nu (t) = \int_0^t \mu (t') dt'
\end{eqnarray}
Using Eqs. (\ref{Eq-3a},\ref{Eq-3b}) we obtain a closed equation for $q$, ie.
\begin{equation}
\label{Eq-4}
\ddot{q}=-\left( \alpha  e^{q-\nu} - \frac{\dot{\alpha}}{\alpha} \right) \dot{q}
\end{equation}
Equation (\ref{Eq-4}) is a unique second order equation that fully captures the dynamics of the SIR infection model. While it is highly nonlinear, it is nevertheless quite useful in determining the infection dynamics since it is general and contains the arbitrary time dependence of both the infection and removal rates. It will be used subsequently in the application of ML techniques to the COVID-19 infection data. In the case of constant infection and removal rates it can be solved exactly; this solution is given in Appendix A.

\section{Time-dependent infection rate equation}

We start with the general Eq. (\ref{Eq-4}) with time dependent infection rate $\alpha (t)$ and  for the sake of simplicity we assume that the recovery rate $\mu (t) \equiv \mu$ is a constant; in this case the expression of Eq (\ref{Eq-4}) simplifies to $\nu =\mu t$ and thus $I(t) = exp \left[ q(t) - \mu t\right]$. While Eq.  (\ref{Eq-4}) is rather involved in several cases we might be interested in the inverse problem where although we know the infection data we are not able to asses directly the applied measures $\alpha (t)$ that generates it. We know, for instance, that a monotonic linear drop in the infection rate, as for instance introduced by gradual social distancing measures results in an approximately Gaussian evolution\cite{BT}. We may thus write Eq. (\ref{Eq-4}) as:

\begin{eqnarray}
\label{Eq-20a}
\frac{d \alpha}{dt} + f(t) \alpha = g(t) \alpha^2 \\
\label{Eq-20b}
f(t) = -\frac{\ddot{q}}{\dot{q}}\\
\label{Eq-20c}
g(t) =  e^{q - \mu t}.
\end{eqnarray}
\\
The Eq. (\ref{Eq-20a}) is a Bernoulli equation\cite{PZ}  that can be turned into a linear first order equation by using the transformation $z(t) = 1/ \alpha (t)$; we obtain
\begin{equation}
\label{Eq-21}
\frac{d z}{dt} - f(t) z + g(t) = 0.
\end{equation}
The general solution thus of Eq. (\ref{Eq-20a}) obtained through the solution of Eq. (\ref{Eq-21}) is
\begin{eqnarray}
\label{Eq-22a}
\frac{1}{\alpha (t)} = C''e^{-F(t) } - e^{-F(t) } \int e^{F(t') } g(t') dt' \\
\label{Eq-22b}
F = - \int f(t') dt',
\end{eqnarray}
where $C''$ is an arbitrary constant.

Let us now consider the case where the infected population behaves similar to a Gaussian 
function \cite{BT}, keeping however also a linear time-term in the exponent that provides some 
time asymmetry, i.e. take 
\begin{equation}
\label{Eq-23}
q(t) = \beta t^2 + \gamma t.
\end{equation}
Simple algebra leads to 

\begin{eqnarray}
\label{Eq-24a}
q-\mu t = \beta t^2 + (\gamma - \mu ) t\\
\label{Eq-24b}
f(t) = - \frac{\ddot{q}}{\dot{q}} =-  \frac{2 \beta}{2 \beta t + \gamma }\\
\label{Eq-24c}
g(t) =  e^{q - \mu t} = e^{\beta t^2 + (\gamma - \mu ) t}\\
\label{Eq-24d}
F = - \int f(t') dt' =  \int \frac{2 \beta}{2 \beta t' + \gamma } dt' =  ln (2 \beta t + \gamma )
\end{eqnarray}
and thus the solution of Eq. (\ref{Eq-22a}) becomes
\begin{eqnarray}
\label{Eq-25a}
\frac{1}{\alpha (t)} = \frac{ \alpha (0) \gamma }{ 2 \beta t + \gamma } +\frac{K(t)}{ 2 \beta t + \gamma }\\
\label{Eq-25b}
K(t) = -
 \int_0^t 
(2 \beta t' + \gamma)e^{\beta {t'}^2 + (\gamma - \mu ) t'}dt' = \left[ 2 \beta -(\gamma - \mu ) 
(\gamma + 2 \beta t) \right] e^{\beta t^2 +(\gamma - \mu ) t} 
\end{eqnarray}
Finally,
\begin{equation}
\label{Eq-26}
\alpha (t) = (2 \beta t + \gamma ) \left [ \alpha (0) \gamma + \frac{2 \beta}{2 \beta t + \gamma } 
- (\gamma - \mu ) e^{\beta t^2 + (\gamma - \mu )t }\right]
\end{equation}

We note that the dominant term is that of  linear decay since at longer times and $\beta < 0$ the Gaussian term in Eq. (\ref{Eq-26}) essentially disappears while the exponential term also decays when $\mu > \gamma$.   In general, of course, the functional dependence of $\alpha (t)$ is more complex and in cases with strong asymmetry introduced by $\gamma$ we have distinctly nonlinear decay.  We observe thus how significant is the precise functional form of the time dependent infection rates for the general evolution of the SIR modeling of the infection phenomenon.

\section{Feature extraction through pre-trained, physics informed neural networks}

The exact mathematical analysis of the SIR model is important for the analysis of the data through ML techniques. The COVID-19 ``first wave" started at different times in various countries and had a completely different evolution.  In countries where very restrictive measures similar to those of China were imposed an effective spreading control swiftly was accomplished. Other countries that either delayed or imposed partial or essentially no measures saw larger numbers in the infected population and slower decay in the infected numbers. Here we make no judgment as to whether measures were ``good" or ``bad", but we simply want to be able to extract the presence of the measures from the dynamics of the infected population. Specifically, we would like to see what is the imprint of social distancing in the distribution of the infected population across the eight model countries we follow.  To accomplish this, we use a strategy that utilizes methods from Artificial Intelligence and in particular Machine Learning (ML). The basic assumption in our approach is that the SIR model can capture the essentials of the epidemic in each country. A direct consequence of this assumption is that we can use simulated data from the SIR model to pre-train the specific neural networks we use for each country.  The specifics of the application of ML in this problem are detailed in the Appendix.   

The application of ML techniques to data often suffers from the fact that data are considered ``pure" with no connection to a specific phenomenon.  A remedy towards introducing specificity is through the use of physics-informed techniques where the ML processes, typically those involving Artificial Neural Networks (ANN), are restricted imposed by physical laws in mathematical form\cite{RPK}. In the specific problem, the SIR equations play this role and put strict bounds to the ANN used for simulating the phenomenon.  Once we have a physics-informed network that is trained on the infected data of the specific country, we may use it to extract the presence and persistence of the social distancing measures typified through the function $\alpha (t)$.  Since the ANN finds a general decay and also given the discussion in the previous section, we posit a linear dependence in the form $\alpha (t) = \sigma_0 + \sigma t $, where the intercept $\sigma_0$ and the slope $\sigma$ are determined through the ANN. The slope $\sigma$, in particular, is a parameter with an important physical significance, since it estimates within the linearized model the degree of efficiency of social distancing.  In other words,  a large value in $\sigma$  describes in an average way a country that followed through the first wave strict social distancing measures while, on the contrary, small values in the slope denote much looser adherence to measures.  We note that these measures are not necessarily the externally imposed ones but also include the self-imposed measures.

Specifically, in this work, we approximate each country's daily reported cases using a deep neural network containing one input node, five hidden layers of 100 nodes with a ``sigmoid" activation function, and one output node. Initially, we use simulated SIR data and an arbitrary linear function for $\alpha(t)$ with a constant value for $\mu$, to train the model. The model is trained, using a custom training loop, by minimizing the mean squared error loss on the data, $MSE_{D}$:

\begin{equation}
MSE_{D} = \frac{1}{N_D} \sum_{i=1}^{N_D} |x_i - \tilde{x}_i|^2,
\end{equation}
where ${ \{ x_i, \tilde{x}_i \} }_{i=1}^{N_D} $ denote the set of the reported, $x_i = ln(I_i)$, and corresponding predicted cases, $\tilde{x}_i = model(t_i)$ and the mean squared error loss defined by Eq.(\ref{Eq-4}) with $\alpha = \alpha(t) = \sigma_0 + \sigma t$ and $\nu = \mu t$ (ie. $\mu$ = constant), $MSE_{SIR}$:
\begin{equation}
MSE_{SIR} = \frac{1}{N_{SIR}} \sum_{j=1}^{N_{SIR}} |f(t_j, \tilde{x}_j, \dot{\tilde{x}}_j, \ddot{\tilde{x}}_j, \sigma_0, \sigma, \mu)|^2,
\end{equation}
where,
\begin{equation}
f(t_j, \tilde{x}_j, \dot{\tilde{x}}_j, \ddot{\tilde{x}}_j, \sigma_0, \sigma, \mu) = \ddot{\tilde{x}}_j + \left( \alpha_j e^{\tilde{x}_j} - \frac{\dot{\alpha}_j}{\alpha_j} \right) (\dot{\tilde{x}}_j + \mu) = 0.
\end{equation}

Then for each of the countries into consideration, we load the real data and smooth it using a seven time-steps moving average. We then scale the data using Min-Max normalization. We load the pre-trained model and allow all its weights to be tuned by minimizing again both the $MSE_{D}$ and $MSE_{SIR}$ loss functions on the country's data, getting at the same time the optimal values for $\alpha(t)$ and $\mu$ for the given country. The pre-trained model is used to accelerate each country's training process. The training process of each country stops using early stopping with a horizon of 200 epochs. 

\begin{figure} [h]
 \includegraphics[width=8.5cm]{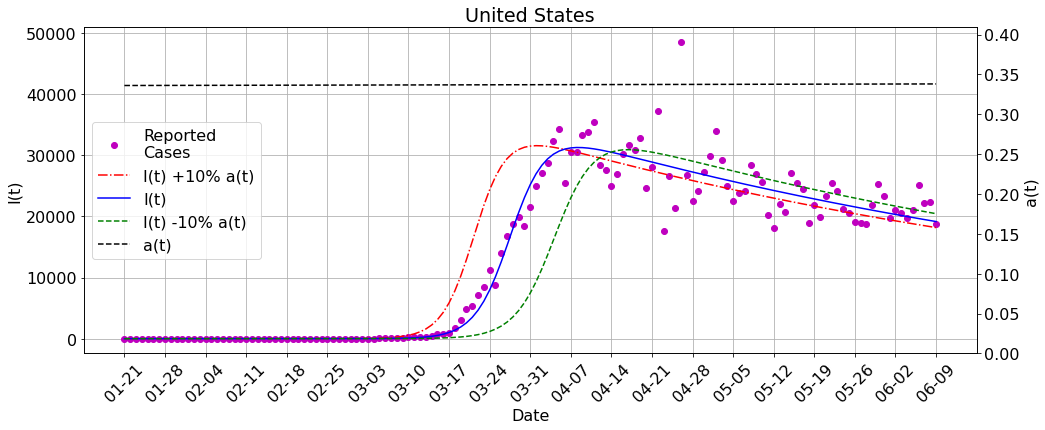} \includegraphics[width=8.5cm]{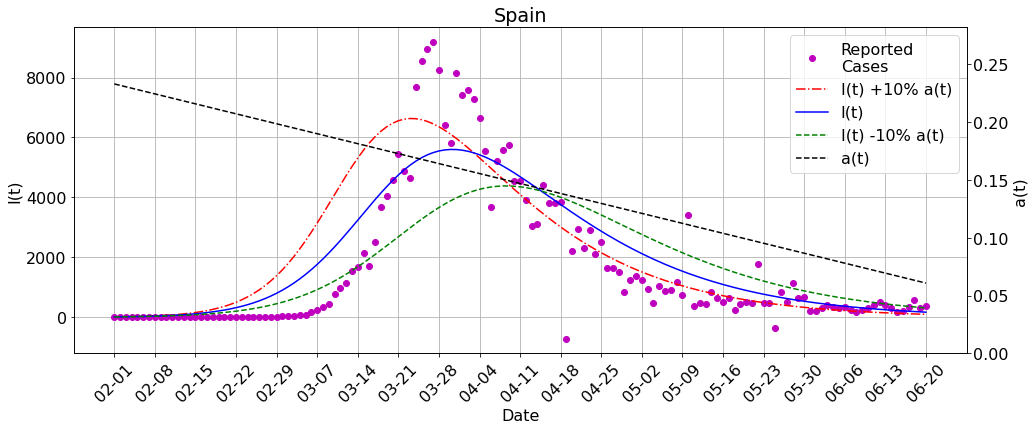} 
 \includegraphics[width=8.5cm]{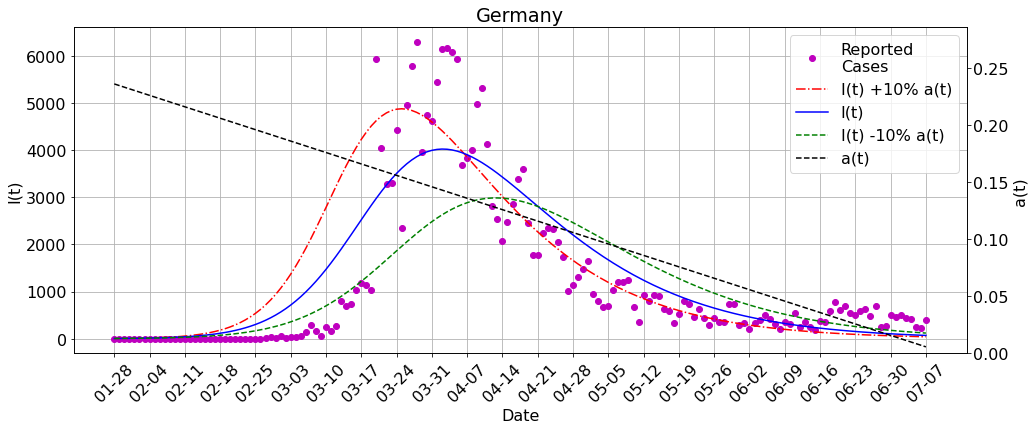}\includegraphics[width=8.5cm]{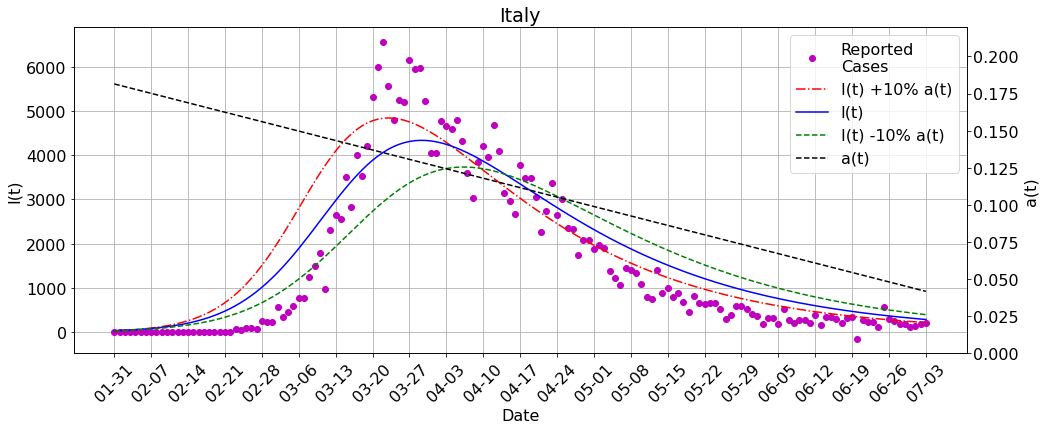}
 \includegraphics[width=8.5cm]{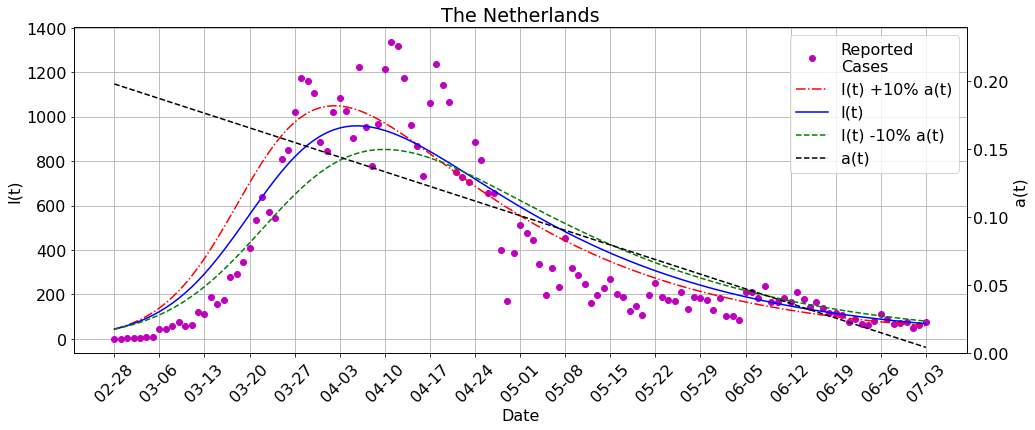} \includegraphics[width=8.5cm]{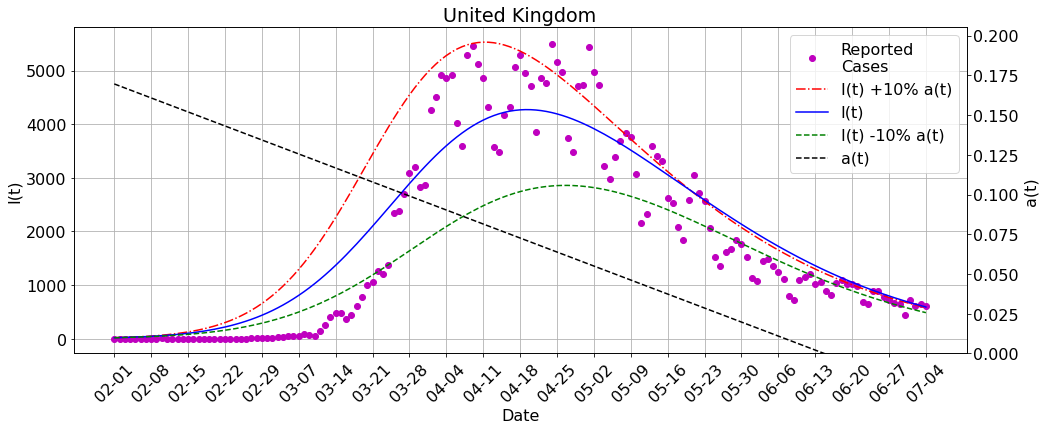}     
 \includegraphics[width=8.5cm]{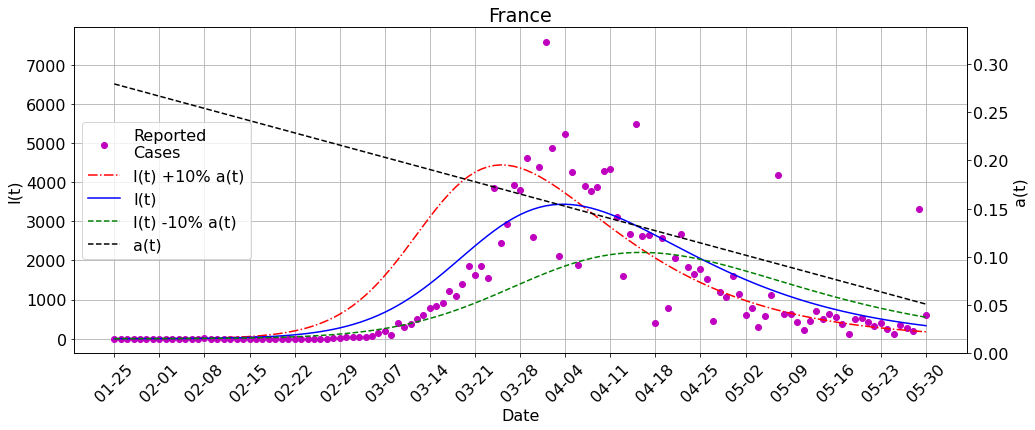} \includegraphics[width=8.5cm]{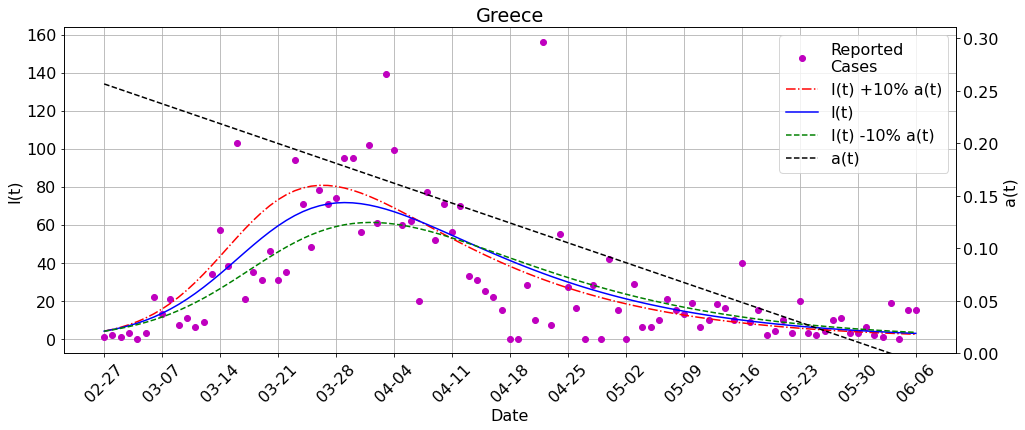}
\caption{Country level predicted infections, $I(t)$, for the extracted $\alpha(t)$ of each country. Magenta dots represent the reported cases of each country. Red, dashed and dotted line, blue solid and green dashed line represent the infections with +10\%, no change, -10\% to the infection rate $\alpha(t)$, respectively. Black dashed line represents the extracted infection rate of each country.}
 \label{Fig1}
\end{figure}

\begin{figure} [h]
 \includegraphics[width=8.5cm]{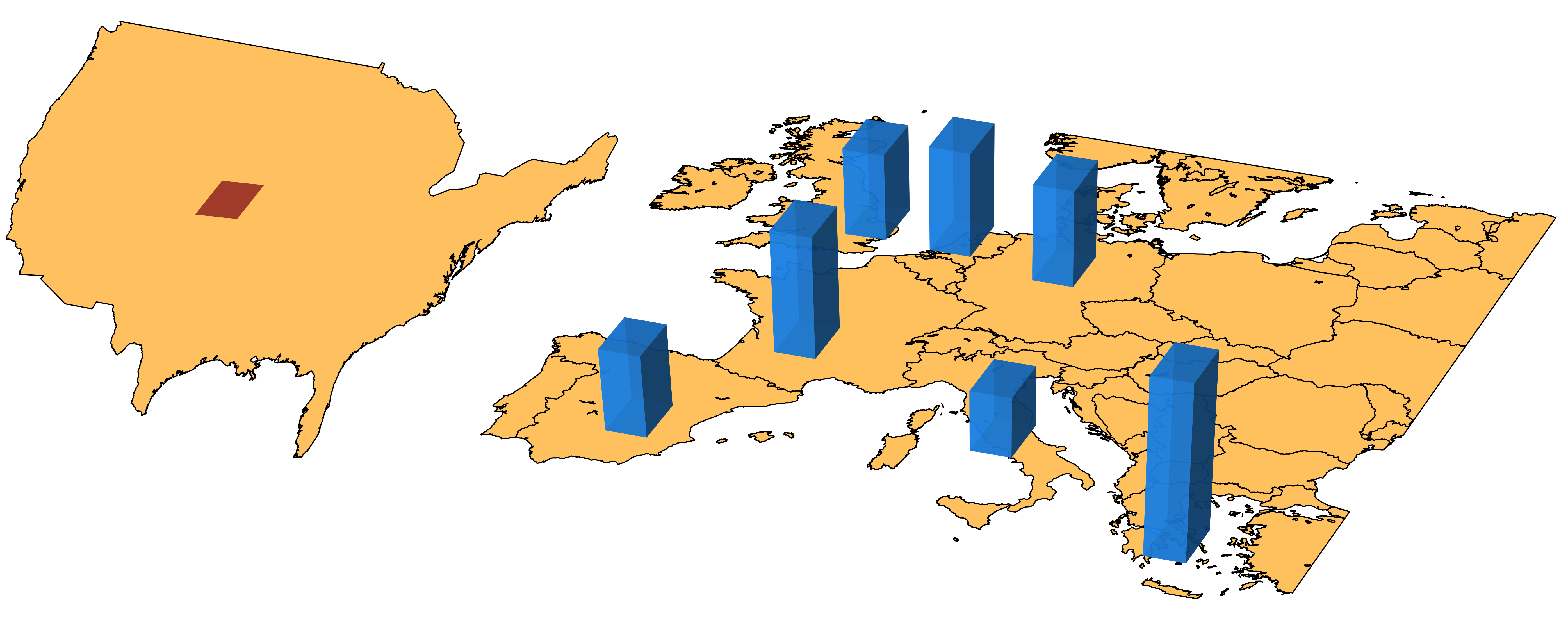} 
\caption{Presentation of the values $\sigma$ derived for the first phase of the infection for each country on a map for better visualization. The red color in the US denotes opposite sign, i.e. in terms of the present interpretation non-adherence to measures on average. In the European countries the value of Greece is maximal.}
 \label{Fig2}
\end{figure}

Having extracted the optimal $\alpha(t)$ as well as $\mu$ for each country, we use them to solve the SIR model, Eqs. (\ref{Eq-1a}, \ref{Eq-1b}). The solution is then fitted to the country's real data using the initial conditions ($I_0$, $S_0$) as fitting parameters. The total number of the predicted cases during the ``first wave" period of each country, including the relative error to the corresponding total number of reported cases and the total number of cases obtained by varying $\alpha(t)$ by $\pm$10\%, is presented in the Table (\ref{Tab1}). A plot of the results for each country is shown in Fig. (\ref{Fig1}). In the map of Fig. (\ref{Fig2}) we portray the results of Table (\ref{Tab1}) in a more graphic way.   

The machine learning algorithms were implemented in Python using TensorFlow/Keras \cite{TFK} and the ADAM \cite{ADAM} optimizer. The data used in this study were published online at OurWorldInData.org\cite{SRC}.

\begin{table}[h!]
\begin{minipage}[t]{0.65\linewidth}
\centering
\begin{tabular}{|l|r|r|r|r|r|c|}
\hline
Country & \multicolumn{2}{|c|}{Total cases} & Error & $\alpha(t)$ + 10\% & $\alpha(t)$ - 10\% & $R^2$\\
			  & Reported & Predicted 				& (\%) & (\% Difference) & (\% Difference) & \\
\hline
USA & 1961185 & 1945830 & -0.78 & 1793214 (-7.84) & 2063029 (6.02) & 0.944\\
\hline
Italy & 240961 & 275667 & 14.40 & 259349 (-5.92) & 284978 (3.38) & 0.863\\
\hline
Spain & 245938 & 280859 & 14.20 & 249833 (-11.05) & 298190 (6.17) & 0.791\\
\hline
UK & 286141 & 312211 & 9.11 & 217528 (-30.33) & 382591 (22.54) & 0.872 \\
\hline
Germany & 186839 & 215563 & 9.45 & 182532 (-15.32) & 233021 (8.10) & 0.776\\
\hline
The Netherlands & 50412 & 55040 & 9.18 & 52524 (-4.57) & 56583 (2.80) & 0.837 \\
\hline
France & 149668 & 163580 & 9.30 & 117539 (-28.15) & 187154 (14.41) & 0.702 \\
\hline
Greece & 2967 & 3014 & 1.60 & 2794 (-7.32) & 3155 (4.68) & 0.540 \\
\hline
\end{tabular}
\end{minipage}\hfill
\begin{minipage}[c]{0.35\linewidth}
\centering
\includegraphics[width = 5cm]{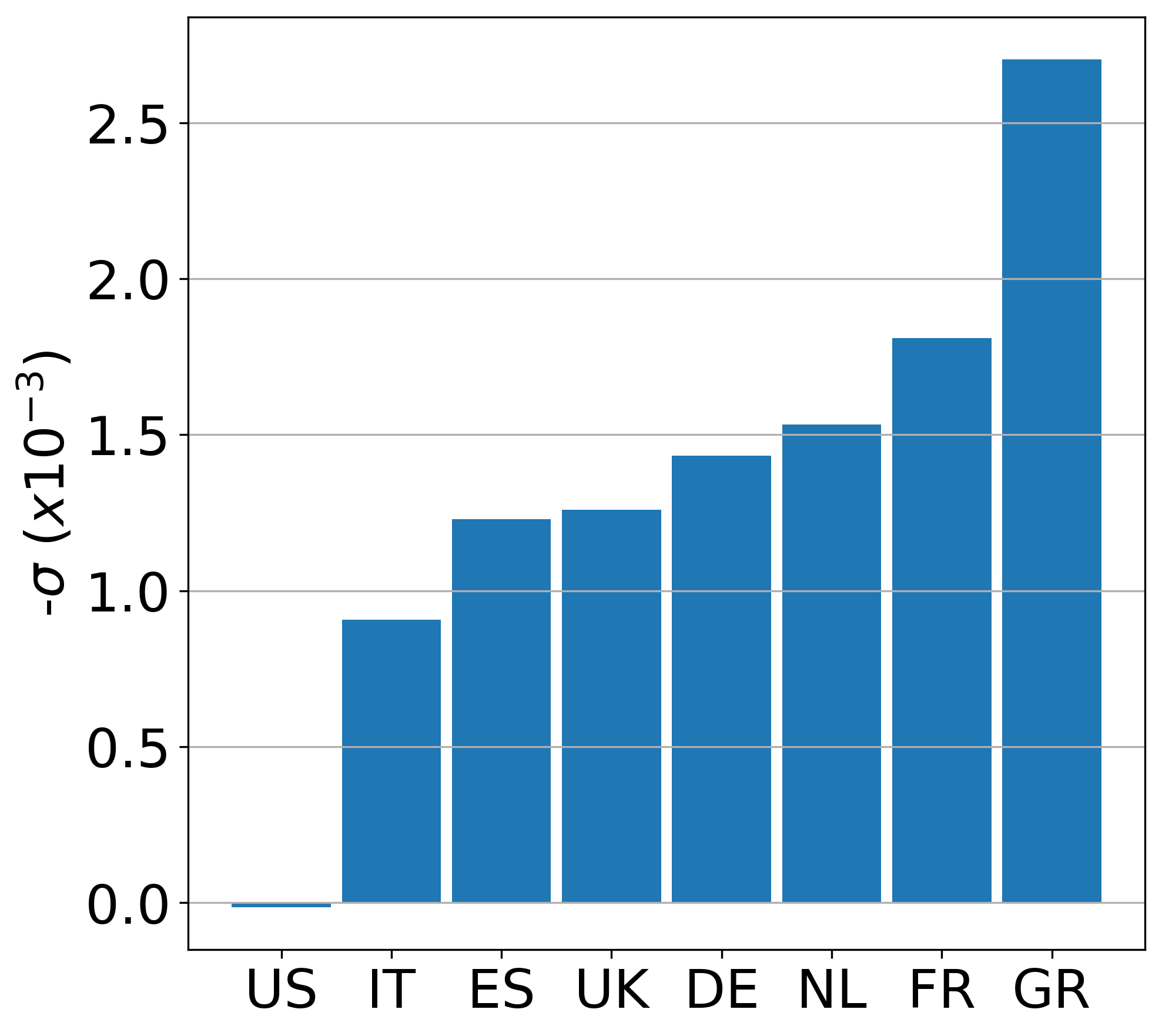}
\end{minipage}
\label{Tab1}
\caption{Left: Total number of reported cases during the ``first wave" for each country and the corresponding predicted cases and percentage error obtained from our model, including the predictions with $\pm$ 10\% variation of a(t). Right: A bar plot of the slope  $\sigma$ of each country signifying the degree of adherence to measures.}
\end{table}

\section{Short term predictions}

The arsenal of physics informed machine learning was used in the previous section in order to extract dynamical parameters such as the time-dependent infection we well as the removal rates from the documented infection data. The procedure through SIR pre-training proved to be quite efficient and gave a hierarchy of $\alpha (t)$ for different countries for the initial period of the infection.  It is both tempting as well as challenging to apply this procedure to the present phase of the COVID-19 pandemic and attempt to make future predictions.  In the process of future point evaluations, our procedure needs to satisfy two constraints; one is the overall mean square minimization that reduces the overall error.  The second is the one imposed by physics, i.e. it must follow the SIR model. In order to accomplish the latter, the procedure needs to know the functional of $\alpha (t)$ as well as the value of $\mu$ at the future points.  We provide this information through the extrapolation of the values from the previous times.  Once these values are known, the SIR dynamics is warrantied and provides the second, physics derive constraint.  

\begin{figure} [h]
\includegraphics[width=8.5cm]{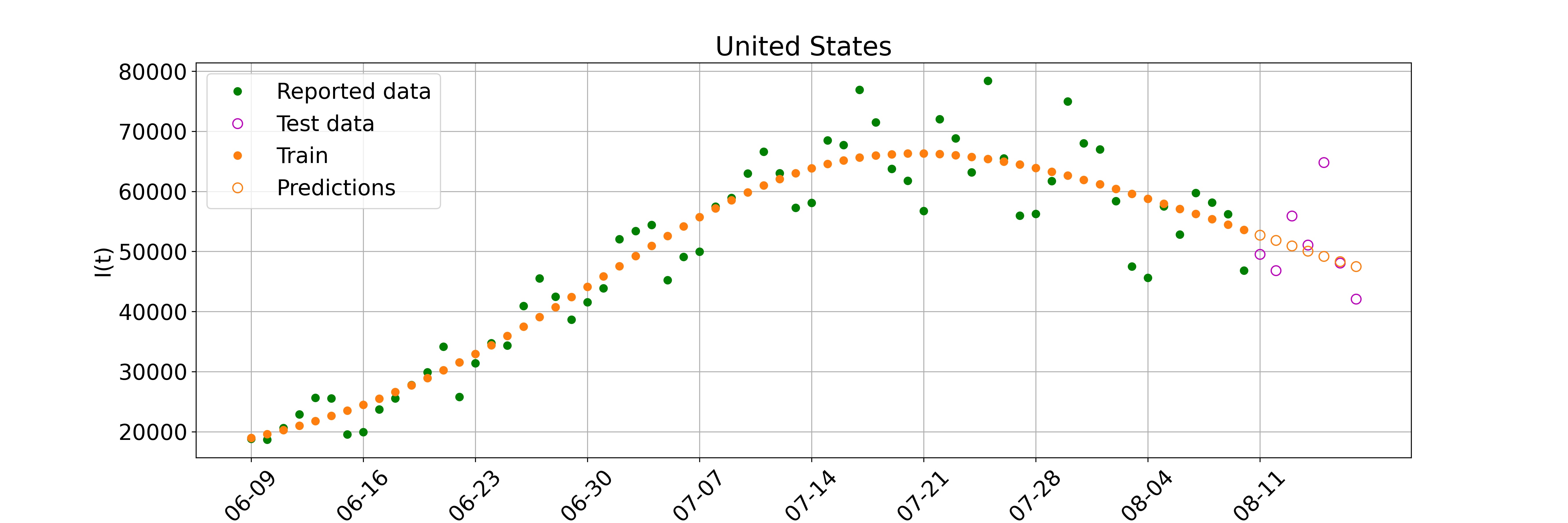} \includegraphics[width=8.5cm]{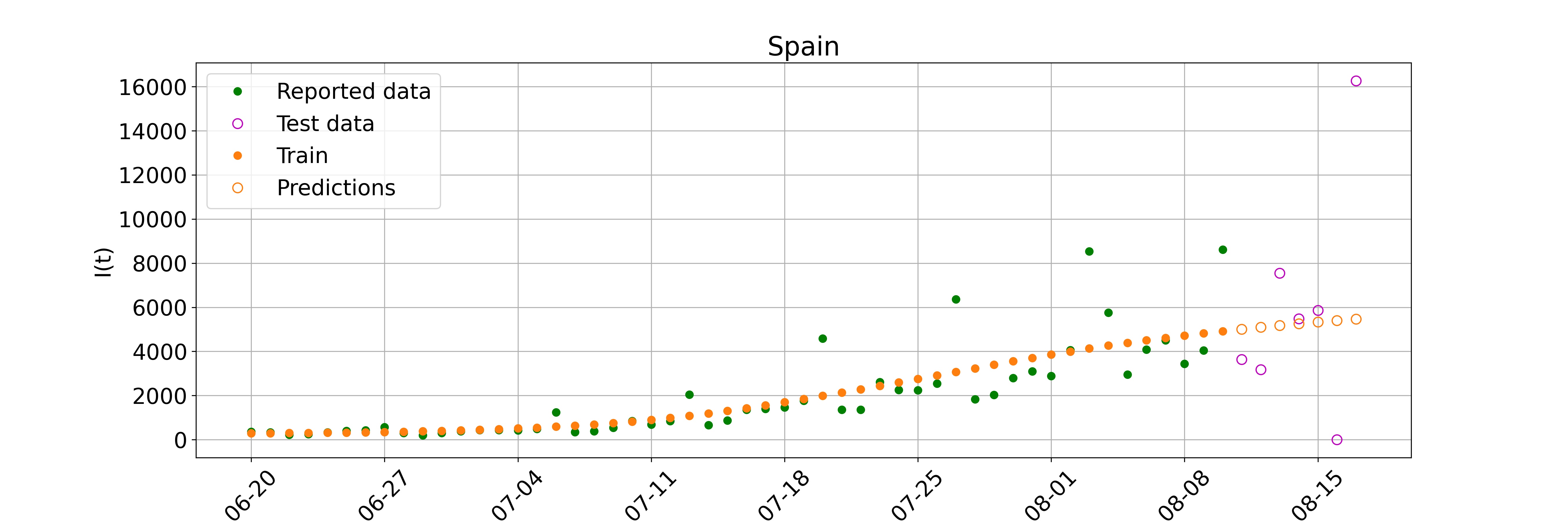} 
\includegraphics[width=8.5cm]{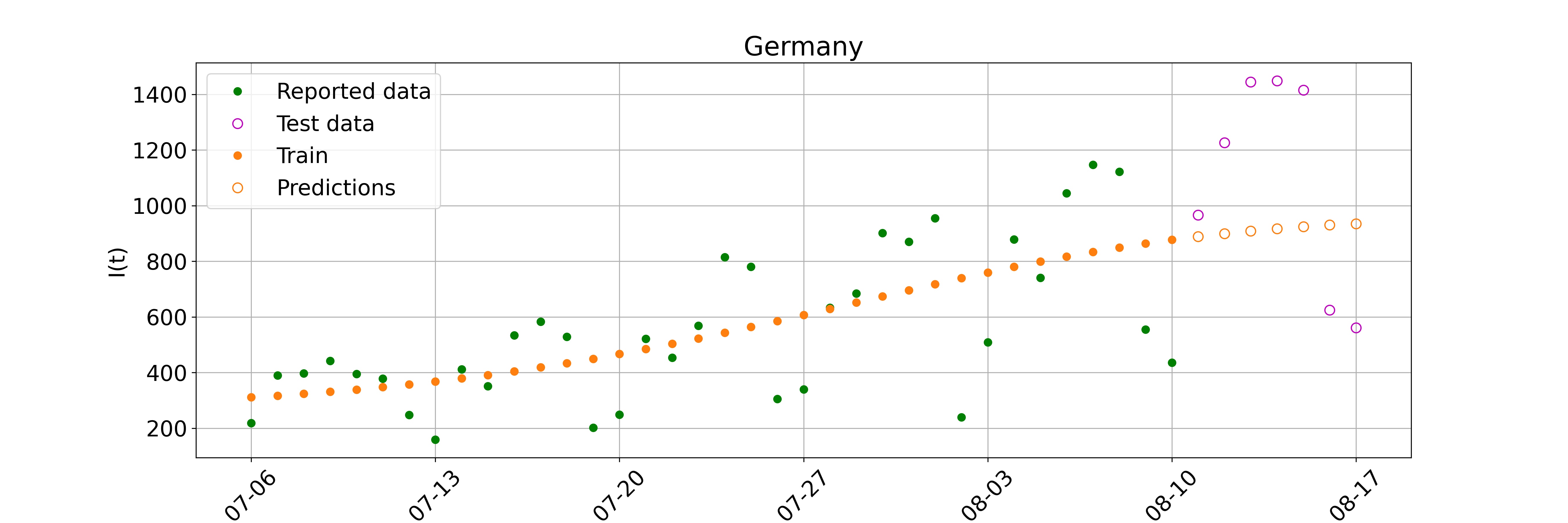} \includegraphics[width=8.5cm]{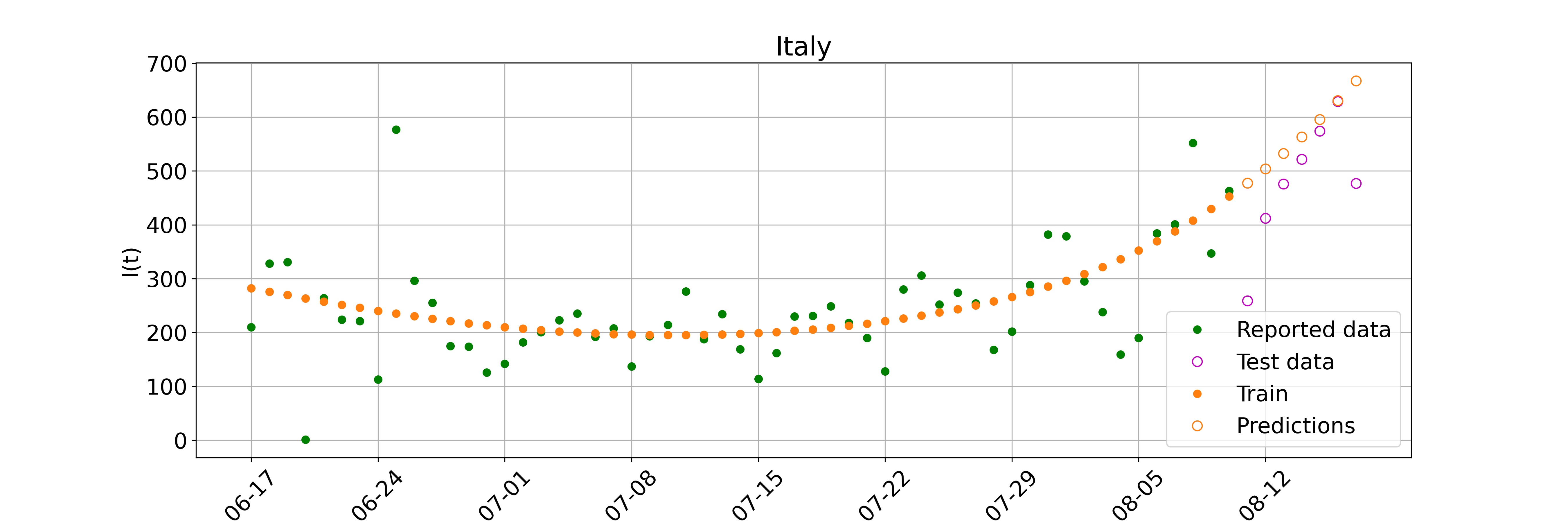}
\includegraphics[width=8.5cm]{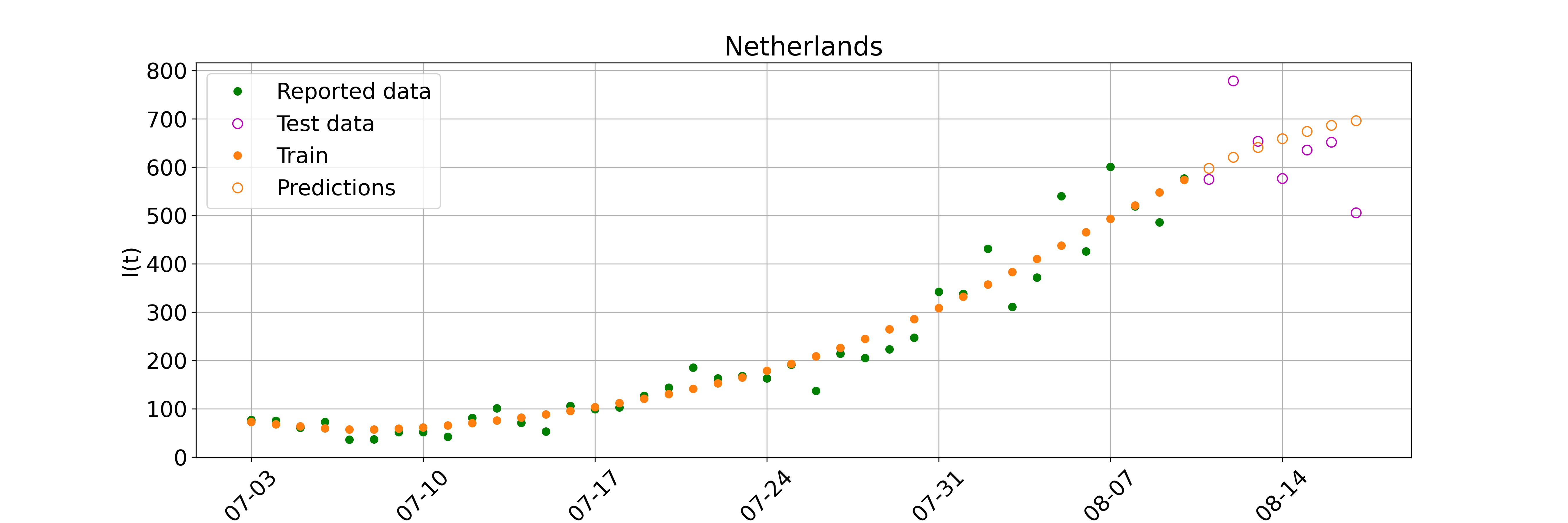} \includegraphics[width=8.5cm]{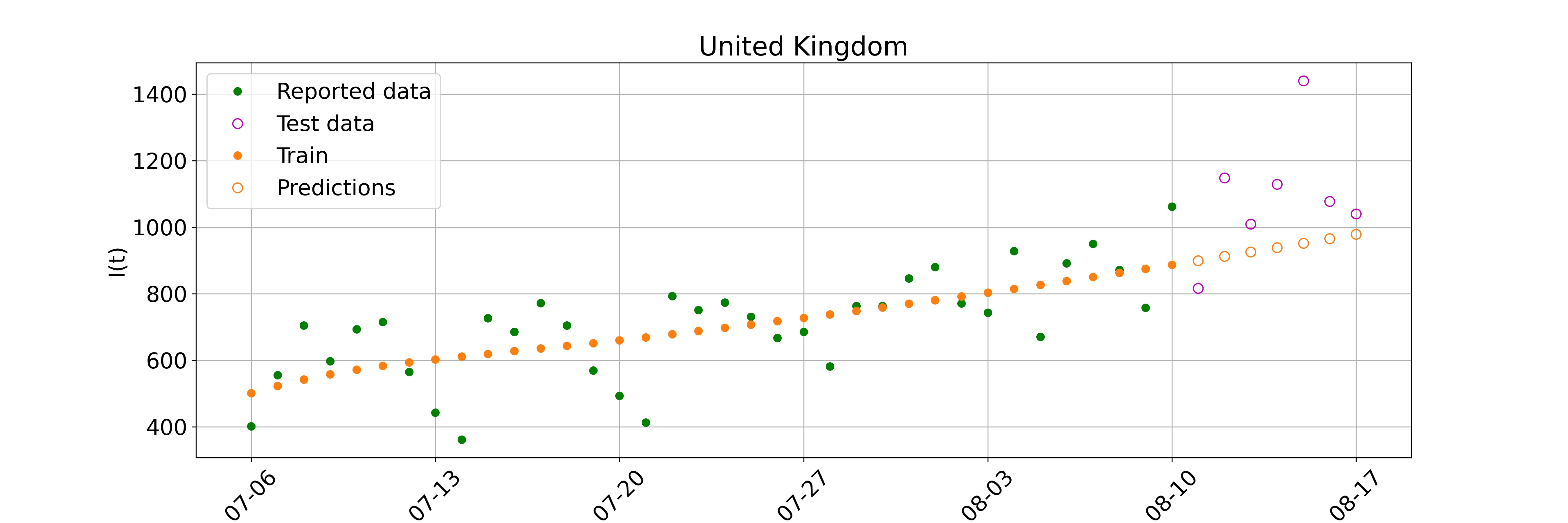}     
\includegraphics[width=8.5cm]{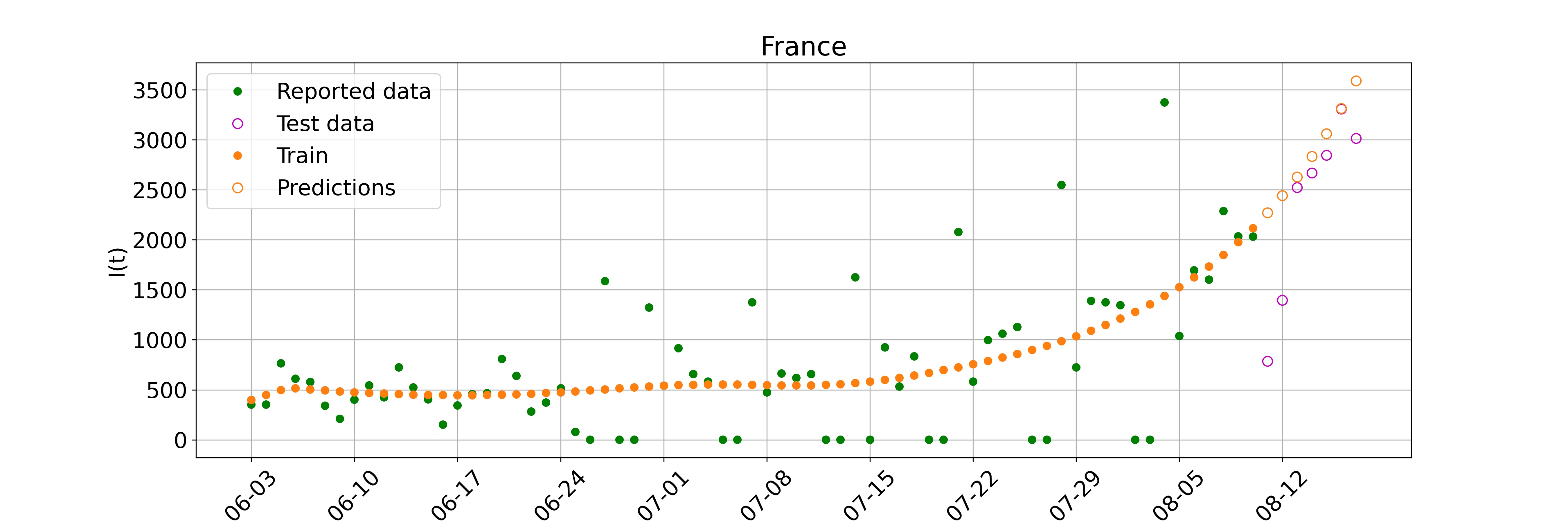} \includegraphics[width=8.5cm]{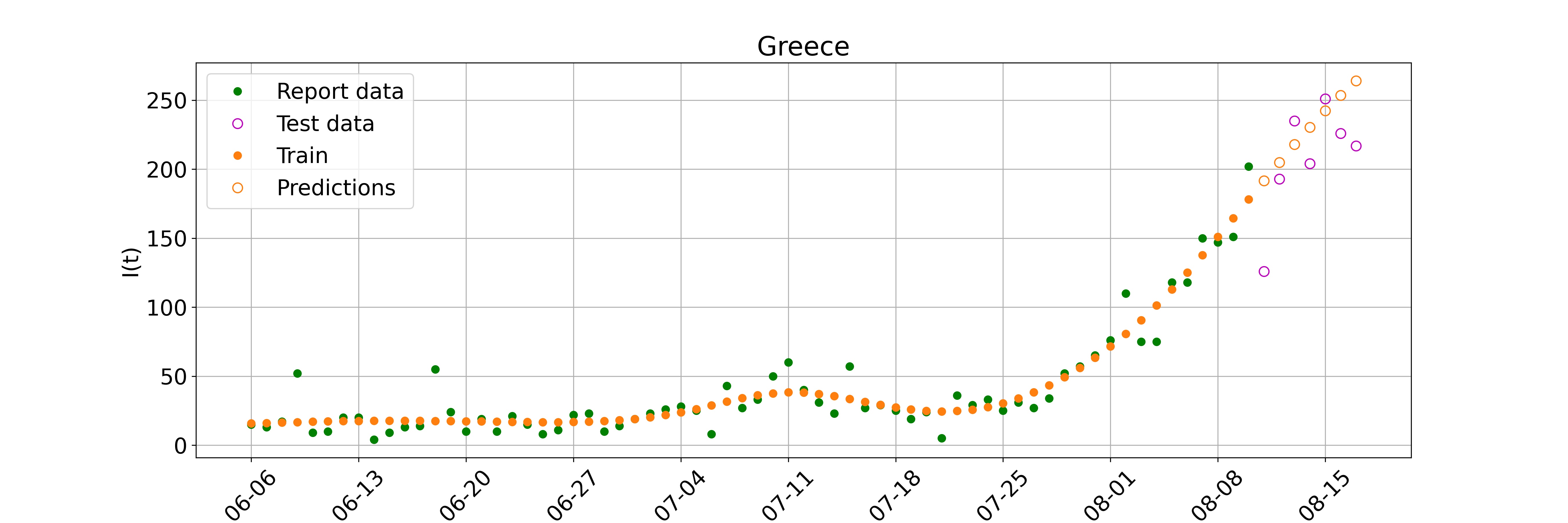}

\caption{PINN short term predictions and comparison with existing data. We use data of the second phase of COVID-19 spreading except for the last week, we train the network and predict the evolution during the last week. In most cases the short term prediction is reasonably good. }
 \label{Fig3}
\end{figure}

In Fig. (\ref{Fig3}) we show the evolution of the current phase of the pandemic as well as the prediction obtained for a horizon of one week.  In preparing these results, we used the available COVID-19 data, starting precisely where the first phase ended and used it up to one week before end dates for all eight countries for  PINN training.  Subsequently, we used the network for prediction and compared the results with the existing data.  We note that the network's short-term predictive power is quite good on average in most countries. 

\section{Conclusions}
The spreading of COVID-19 has generated a wave of illness and death worldwide, accompanied by a severe disruption in financial, educational, commercial activities, global travel etc \cite{RM,KAN,JS}. During the first phase of the spreading, there were different approaches to the measures to be taken to slow it down.  Different countries reacted in different ways, and, as a result, the epidemic dynamics proceeded differently. The infection curves were different and dependent strongly both on the imposed social distancing measures and the adoption of responsible practices from individuals. One important aspect of the pandemic is to find ways to assess the degree to which the social distancing measures where followed. It is not trivial to extract this information from the data since the infection dynamics are directly related to the imposed social distancing measures.  Furthermore, this cannot be done in a completely model free way, and thus assumptions about both the model and the way measures are imposed are important.

In the present work, we followed an early attempt \cite{BT} and used the publicly available infection data to assess the effectiveness and adherence to social distancing in different countries; in doing this analysis, we assumed that the mathematical model underlying the infection dynamics is the simplest SIR model. Before using the arsenal of ML we tackled the model analytically and produced two basic results; the first one is a general analytical solution for the model obtained through a specific exponential ansatz.  The second, also dependent on this ansatz, is a differential equation for the function $\alpha (t)$ that describes the time dependent nature of the infection rate. The latter depends strictly on the imposed social distancing measures as well as the practices of the individuals.  We pointed out, as also in reference \cite{BT} , that a linear drop in the infection rate leads to an approximate Gaussian functional dependence in the infected population. The specifics of this functional form depends both on the form as well as values of  $\alpha (t)$ but also on the removal rate $\mu$. 

In order to extract the time-dependent infection rate from the data, we used physics-informed neural networks, i.e. a machine learning method that uses input from the actual model assumed, viz. SIR. This input,  together with the real infection data from each country we considered,  led to a prediction of the assumed linear in time infection rate. The data derived slope $\sigma$ signifies the adherence of each country to social distancing.  In Greece, for instance, the slope is large in absolute value, designating strong application of the imposed measures by the individuals.  In the other extreme, we find the USA with a practically zero slope, demonstrating that the measures taken had low efficiency.
 The other six countries we analyzed fall in intermediate locations between these two extremes. Application to the SIR model of each country, an alternative infection rate that differs by a few percent ($\pm 10\%$) in total from the one obtained through ML gives an estimate of how dependent the infection is on the applied measures.  We find that this variation, while it affects the early SIR fast rise strongly, results in quite a different infection decay and horizon in countries like the UK. 

Once we know how the PINN behaves with the data fort the initial period of the infection, we may use it for the second phase.  We consider that the latter starts from the end of the initial period and reaches the present day.  Thus, we use country infection data during this period except for the last week to train the network and subsequently make predictions for the last week and compare it with real data.  We find that while the short term predictive power of PINN is good, it has large deviations in countries where the data appear to have a rather stochastic character. 

The basic conclusion of this work is that the use of physics-informed ML may enable the extraction of COVID-19 infection information in different countries, show how different measures and practices are directly reflected in the data and ultimately make predictions.  The use of physics in machine learning gives specificity to the data, but, on the other hand, is restricted and some times limited to inserted physics knowledge. The present approach assumes a well-mixed, essentially uniform country, an assumption that is introduced through the use of the SIR model.  However, countries have regions, and each region may behave differently for geographical, environmental, cultural, as well as population reasons.  If regional data is available, one can go one step further and introduce spatial in addition to temporal distribution in the infection and from this be able to obtain more accurate results and predictions. We believe the methodology used in this work may be extended in this more realistic case and provide a more direct approach to local dynamics and the effectiveness of imposed measures at a local level.

\section*{Appendix A: Exact solution of the SIR model for time-independent infection rates}
\subsection{Differential Equations}
For the simpler case of constant $\alpha$ and $\mu$, Eq. (\ref{Eq-4}) becomes
\begin{equation}
\label{Eq-5}
\ddot{q}=- \alpha  e^{q-\mu t}  \dot{q}
\end{equation}
Introducing the transformation $x=q-\mu t$ we turn Eq. (\ref{Eq-5}) into the following form:
\begin{equation}
\label{Eq-6}
\ddot{x} + \alpha  e^{x}\dot{x}+ \alpha \mu e^{x}  =0
\end{equation}
The new initial conditions are $x(0) = ln I(0)$ and $\dot{x}(0) = aS(0) - \mu$.
The Eq. (\ref{Eq-6}) is a Lienard Equation that can be turned into an Abel equation  through the introduction of the transformation \cite{Polyanin}
\begin{equation}
\label{Eq-7}
y(x) = \dot{x}
\end{equation}
We obtain the following Abel equation of the second kind:
\begin{eqnarray}
\label{Eq-8a}
y y_x = f_1 (x) y + f_0 (x)\\
\label{Eq-8b}
f_1 (x) = - \alpha e^x ~,~f_0 (x) =-  \alpha \mu e^x
\end{eqnarray}
We introduce further the variable $\xi$ as follows
\begin{equation}
\label{Eq-9}
\xi = \int f_1 (x) dx =  - \alpha e^x
\end{equation}
Since $y_x = y_{\xi}  f_1 (x)$, Eq. (\ref{Eq-8a}) becomes
\begin{equation}
y y_{\xi} = y +  \mu
\label{Eq-10}
\end{equation}
The Eq. (\ref{Eq-10}) has the implicit solution
\begin{equation}
\xi = y -  \mu ln|y+ \mu | +C
\label{Eq-11}
\end{equation}
where $C$ is an arbitrary constant, or
\begin{equation}
-\alpha e^x  = y -  \mu ln|y+\mu | +C
\label{Eq-12}
\end{equation}
Once the  solution $y = y (x) $ is substituted to Eq. (\ref{Eq-10}) in the form
\begin{equation}
\label{Eq-13}
t= \int^x  \frac{dx}{y(x)}
\end{equation}
we have the implicit solution $t= t(x)$ for Eq. (\ref{Eq-9}). Upon inversion of this solution we may obtain $q(t)$ and thus have a solution for the original SIR equation.

\subsection{Initial Conditions}
The Eq. (\ref{Eq-5}) is a second-order equation while the SIR system of Eq. (\ref{Eq-1a},\ref{Eq-1b}) constitutes a system of two first-order equations with initial conditions $S(0)$ and $I(0)$. It is easy to see that $q(0) = ln I(0)$ while $\dot{q}(0) = \alpha S(0)$. Thus for Eq. (\ref{Eq-9}) we have the following initial conditions $x(0) = q(0) = ln I(0)$ and $\dot{x}(0) = \dot{q}(0) -\mu = \alpha S(0) - \mu$.  Since both susceptible and infected variables are percentages over the total population, the range of the $q=q(t)$ variable is $(-\infty, 0]$ while $x(t)$ takes similarly values in the same range.  

Let us designate for simplicity the values at $t=0$ of $x(0) = k$ and $\dot{x}(0) = m$; clearly $k = ln I(0) < 0$ and $m = \alpha S(0) - \mu$. The latter can be either positive, negative or zero, depending on the initial state of the infection and the corresponding infection rate $R_0$.  The defining transformation of Eq. (\ref{Eq-7}) at $t=0$ becomes $y(k) = m$ and thus the constant $C$ in Eq. (\ref{Eq-12}) is
\begin{equation}
C= \alpha e^k  - m -+ \mu ln|m+\mu | \equiv -  \alpha I(0) - \alpha S(0) +\mu + \mu ln | \alpha S(0) |
\label{Eq-14}
\end{equation}

In other words, the solution Eq. (\ref{Eq-12}) of the differential equation of Eq. (\ref{Eq-10}) should be solved for $x \ge k$ since at $t=0$ we have $x(0) = k$. Thus, the original SIR problem has solution given by 
\begin{equation}
- \alpha e^x  = y -  \mu ln|y+\mu | - \alpha I(0) - \alpha S(0) +\mu +  \mu ln | \alpha S(0) |
\label{Eq-15}
\end{equation}
or, in the equivalent form
\begin{equation}
ln\left[\frac{e^y }{|y+\mu |^\mu }\right] = - \alpha e^x  +\alpha I(0) + \alpha S(0) -\mu - \mu ln | \alpha S(0) |
\label{Eq-15a}
\end{equation}
for $x \ge lnI(0)$ and with the final implicit formula given through the integral of Eq. (\ref{Eq-13}) modified as follows:
\begin{equation}
\label{Eq-16}
t= \int_k^x  \frac{dx}{y(x)} \equiv \int_{ln I(0)}^{ln I(t)}  \frac{dx}{y(x)} 
\end{equation}
%\begin{figure} [h]
% \includegraphics[width=150mm]{Figures/Carnot.pdf}
 % \caption{The Carnot engine operates between two heat reservoirs and performs useful work $W$.  %The 
 % piston of the engine expands when is in contact with the hot bat and contracts when touches the %cold bath. The diagram shows the flow of heat in and heat out while useful work is performed.}
  %\label{Fig1}
%\end{figure}

\subsection{Approximate solutions}

We can write Eq. (\ref{Eq-6}) in the following form
\begin{equation}
\label{Eq-17}
x'' + \lambda  e^{x} x' +  e^{x}  =0
\end{equation}
where $\lambda = \sqrt{\frac{\alpha}{\mu}}$ and the prime derivative is wrt the new time variable
$\tau = \sqrt{\alpha \mu } t$.

A particular solution of Eq. (\ref{Eq-17}) is given by  $x(\tau ) = - \frac{\tau}{\lambda}$, or $x(t) = - \mu t$.  This is, of course, a trivial solution since it corresponds to the solution $q(t) = 0$ of Eq. (\ref{Eq-5}) and thus to a constant infection population.

To find approximate solutions we may look into the limits $\lambda >>1$ and $\lambda << 1$. In the former case, we may ignore the last term of Eq. (\ref{Eq-17}) and solve the equation
\begin{equation}
\label{Eq-18}
x'' + \lambda  e^{x} x'  =0
\end{equation}
Its solution in the original variables is
\begin{eqnarray}
\label{Eq-18a}
t  = \int \frac{dx}{C - \alpha e^x}  = \frac{x}{C} - \frac{ln (C -  \alpha e^x ) }{C~ ln ~e}\\
\label{Eq-18b}
C= \alpha S(0) - \mu + \alpha I(0)
\end{eqnarray}
In the second case we ignore in turn the first derivative term of Eq. (\ref{Eq-17}), i.e.
\begin{equation}
\label{Eq-19}
x'' +  e^{x}  =0
\end{equation}
and obtain the solution in the original variables
\begin{eqnarray}
\label{Eq-19a}
t = \int \frac{dx}{\sqrt{2 (C' - \alpha \mu  e^x )} }=- \frac{2 arctanh\left[\frac{\sqrt{C'- \alpha \mu e^x}}{\sqrt{C}}\right]}{\sqrt{C} ln ~e}\\
\label{Eq-19b}
C' = \frac{\left[\alpha S(0) - \mu \right]^2}{2} - \alpha \mu I(0) 
\end{eqnarray}

\section*{Appendix B: Machine learning application procedure}

In this Appendix we describe the methods used in this work through a flowchart. We note that the after the extraction of the infection rate from the data we use the SIR model for further investigation of the infection.\\
\\

\usetikzlibrary{shapes.geometric, arrows, calc}

\tikzstyle{decision} = [diamond, draw, text width=4.5em, text badly centered, node distance=3cm]
\tikzstyle{block} = [rectangle, draw, text width=10em, text badly centered, rounded corners, minimum height=4em]  
\tikzstyle{line} = [draw, -latex']
\tikzstyle{terminator} = [ draw, ellipse, node distance=3cm, minimum height=2em]  

\begin{tikzpicture}[node distance=2cm, auto]  
  \node [block]           										 (cm)  {Create the Model. Data Driven \& PINN.};
  \node [block, right of = cm, xshift=3cm]   (tsd)  {Train the Model on Simulated SIR Data.};
  \node [block, below of = tsd]   (lcrd) {Load Country's \\ Real Data \& the Pre-Trained Model.};
  \node [block, below of = lcrd]  (smooth) {Smooth Country's Data using a seven time-steps moving average.};
  \node [block, below of = smooth]  (train) {Train the Model \\ using Early Stopping.};
  \node [block, below of = train]  (extract) {Extract Country's $x(t)$, $\alpha(t)$ and $\mu$.};
  \node [block, right of = extract, xshift = 3cm]  (sim) {Solve the SIR model \\ using the extracted $\alpha(t)$ and $\mu$.};
  \node [block, above of = sim, yshift = 6.5mm] (fit) {Fit to Country's Real Data using the initial conditions as fitting paramaters.};
  \node [decision, above of = fit, yshift = 3.5mm] (check) {Other Countries into consideration?};
  \node [block, text width=5em, right of = check, xshift = 2cm] (stop) {Stop};

  \path [line] (cm)  -- (tsd);
  \path [line] (tsd)  -- (lcrd);
  \path [line] (lcrd) -- (smooth);
  \path [line] (smooth) -- (train);
  \path [line] (train) -- (extract);
  \path [line] (extract) -- (sim);
  \path [line] (sim) -- (fit);	
  \path [line] (fit) -- (check);

  \path [line] (check) -- node[anchor=south] {Yes} (lcrd.east);
  \path [line] (check.east) -- node {No} (stop.west);
    
\end{tikzpicture}

\end{document}